\documentclass[aps,pre,twocolumn,superscriptaddress,showpacs]{revtex4}
\usepackage{amsmath}
\usepackage{amssymb}
\usepackage{tikz}
\usepackage{graphicx}
\usepackage{dcolumn}
\usepackage{bm}
\usepackage{epsfig}
\usepackage{psfrag}
\def\II{\hbox{$1\hskip -1.2pt\vrule depth 0pt height 1.6ex width 0.7pt\vrule depth 0pt height 0.3pt width 0.12em$}}

\newcommand{\refsec}[1]{\mbox{Sec.~\ref{#1}}}

\newcommand{\T}{${\mathcal T}\,$}

\begin{document}

\newcommand{\IKP}{Institut f{\"u}r Kernphysik, Technische Universit{\"a}t
Darmstadt, 64289 Darmstadt, Germany}
\newcommand{\IIS}{Indian Institute of Science, Bangalore - 560012, India}

\title{Cross-section Fluctuations in Open Microwave Billiards and Quantum Graphs: The Counting-of-Maxima Method Revisited} 

\author{B.~Dietz} \email{dietz@ikp.tu-darmstadt.de}
\affiliation{\IKP}
\author{A.~Richter}
\affiliation{\IKP}
\author{R.~Samajdar}
\affiliation{\IIS}

\date{\today}

\begin{abstract}
The fluctuations exhibited by the cross-sections generated in a compound-nucleus reaction or, more generally, in a quantum-chaotic scattering process, when varying the excitation energy or another external parameter, are characterized by the width $\Gamma_{\rm corr}$ of the cross-section correlation function. In 1963 Brink and Stephen [Phys. Lett. {\bf 5}, 77 (1963)] proposed a method for its determination by simply counting the number of maxima featured by the cross sections as function of the parameter under consideration. They, actually, stated that the product of the average number of maxima per unit energy range and $\Gamma_{\rm corr}$ is constant in the Ercison region of strongly overlapping resonances. We use the analogy between the scattering formalism for compound-nucleus reactions and for microwave resonators to test this method experimentally with unprecedented accuracy using large data sets and propose an analytical description for the regions of isolated and overlapping resonances. 
  
\end{abstract}


\maketitle
\section{\label{Intro} Introduction} The ubiquity of quantum chaotic scattering is well recognized owing to its fingerprints in diverse fields ranging from compound-nucleus scattering \cite{Verbaarschot1985,Guhr1998,Mitchell2010} to electron transport through disordered mesoscopic samples \cite{Beenakker1997,Alhassid2000}. Owing to the quasibound states of an open system that manifest themselves as resonances in the associated scattering processes, the cross-sections evince random fluctuations with respect to the energy $E$ or some other generic  parameter that are characterized in terms of the cross-section correlation functions. According to the statistical theory of compound-nucleus reactions~\cite{Ericson1963} the coherence width $\Gamma_{\rm corr}$ of the fluctuations is determined by the resonance widths. However, an alternative approach to effectively determine it from the number of maxima exhibited by the cross-sections as function of the excitation energy was introduced in Ref.~\cite{Brink1963}. It was proposed that the product of the number of maxima per unit resonance energy, $K_N$, and the width $\Gamma_{\rm corr}$ is constant, $K_N\Gamma_{\rm corr}=0.5\,b_N$, where $N$ is the number of statistically independent channels participating in the reaction and $b_N$ is a constant. Taking into account finite range of data effects~\cite{Dallimore1965} yielded for a system with a large number of reaction channels $b_\infty=2\sqrt{3}/\pi\simeq 1.1$. For two reaction channels a value of $b_2=0.78$ was obtained~\cite{Brink1963}. The dependence of $b_N$ on the experimental energy resolution of the excitation functions was demonstrated in Refs.~\cite{Bizzeti1967,Roeders1968} for a single channel as well as many open channels. The counting-of-maxima method was believed to yield more accurate results than the determination of $\Gamma_{\rm corr}$ from the autocorrelation functions~\cite{Burjan1986}. Applications to numerical examples have been presented in Refs.~\cite{VanDerWoude1966,Bonetti1983}. 

The main ingredient of all these findings was that one can infer the cross-section correlation width associated with a compound-nucleus reaction by simply counting the maxima emerging when plotting the cross-sections versus the excitation energy. This procedure, actually, was proposed at the time~\cite{Brink1963} as an add-on to the standard analysis of the cross-section fluctuations~\cite{Ericson1963} and was restricted to the Ericson region of strongly overlapping resonances. Recently, the counting-of-maxima method was extended to the regions of isolated and overlapping resonances~\cite{Ramos2011,Barbosa2013,Hussein2014}. There, an analytical expression for the dependence of the product of the number of maxima and the width of the associated cross-section correlation function as function of the energy or, more generally, of some parameter, on the tunneling probability was derived for the cases, that the latter has a Lorentzian and a squared Lorentzian shape, respectively. The findings of Refs.~\cite{Ramos2011,Barbosa2013} actually inspired the investigations described in the present article.

Based on the premise that the $S$-matrix formalism for microwave resonators~\cite{Richter1999} is equivalent to that originally derived for compound-nucleus reactions \cite{Mahaux1969}, we examined the relation of the mean density of maxima in the cross-section fluctuations to the correlation width $\Gamma_{\rm corr}$ in chaotic scattering experiments with unprecedented accuracy. The experimental data, obtained from measurements documented in Refs.~\cite{Dietz2009a,Dietz2010} in the regions of isolated and overlapping resonances, illustrates parallels with studies conducted in Refs.~\cite{Lynn1968,Ericson1960} in the context of nuclear physics \cite{Ericson1966, Richter1974,Mitchell2010}. In addition, the effect of time-reversal invariance (\T) violation, which has been tested in nuclear spectra \cite{French1985} and for the Ericson region in compound-nucleus reactions \cite{Ericson1966a, Mahaux1966, Witsch1967, Blanke1983, Boose1986, Harney1990}, is inspected in this regard. From a theoretical perspective, simulations are performed using random matrix theory (RMT) and the Verbaarschot-Weidenm\"uller-Zirnbauer (VWZ) approach~\cite{Verbaarschot1985}, accompanied by a formal appraisal of chaotic scattering on networks of quantum wires~\cite{Londergan1999}, so-called quantum graphs,~\cite{Texier2001,Kottos2003,Pluhar2013,Pluhar2013a,Pluhar2014}. In light of recent studies of the statistical properties of electronic transport in ballistic open quantum dots~\cite{Ramos2011}, we take cognizance of the analogy between universal conductance fluctuations described by Gaussian processes in quantum dots and compound-nucleus fluctuations~\cite{Guhr1998}. This enables us to analyze and extend the main results of Ref.~\cite{Barbosa2013}, which were originally gleaned from the $S$-matrix describing electron flow in quantum dots, and exhibit their applicability in the wider context of microwave billiards, thereby underlining the subtle interconnections between these seemingly unrelated fields. 

We applied the counting-of-maxima method to spectra, that were measured with a flat microwave resonator with the shape of a chaotic tilted-stadium billiard, and at superconducting conditions in a resonator with a parametric shape, respectively. The aim was to verify the applicability of the method in the region of strongly overlapping resonances~\cite{Brink1963,Dallimore1965} and to verify the validity of the analytical results of Refs.~\cite{Ramos2011,Barbosa2013,Hussein2014} experimentally in the regions of isolated and overlapping resonances. Note, that for the largest achieved ratios of the resonance width and spacing the fluctuation properties of the $S$-matrix elements are already well described by the predictions for the Ericson region~\cite{Dietz2010a}. The experimental results are presented in Secs.~\ref{Exp1} and~\ref{Exp2}, respectively. In order to corroborate our findings we performed RMT simulations, that were based on the scattering formalism introduced in~\refsec{RMT}. Furthermore, we analyzed spectra computed for open quantum graphs, as outlined in~\refsec{Graph}. 

\section{\label{Exp1} Fluctuations in an open microwave billiard with and without \T violation} 
\begin{figure}[h]
        \includegraphics[width=0.8\linewidth]{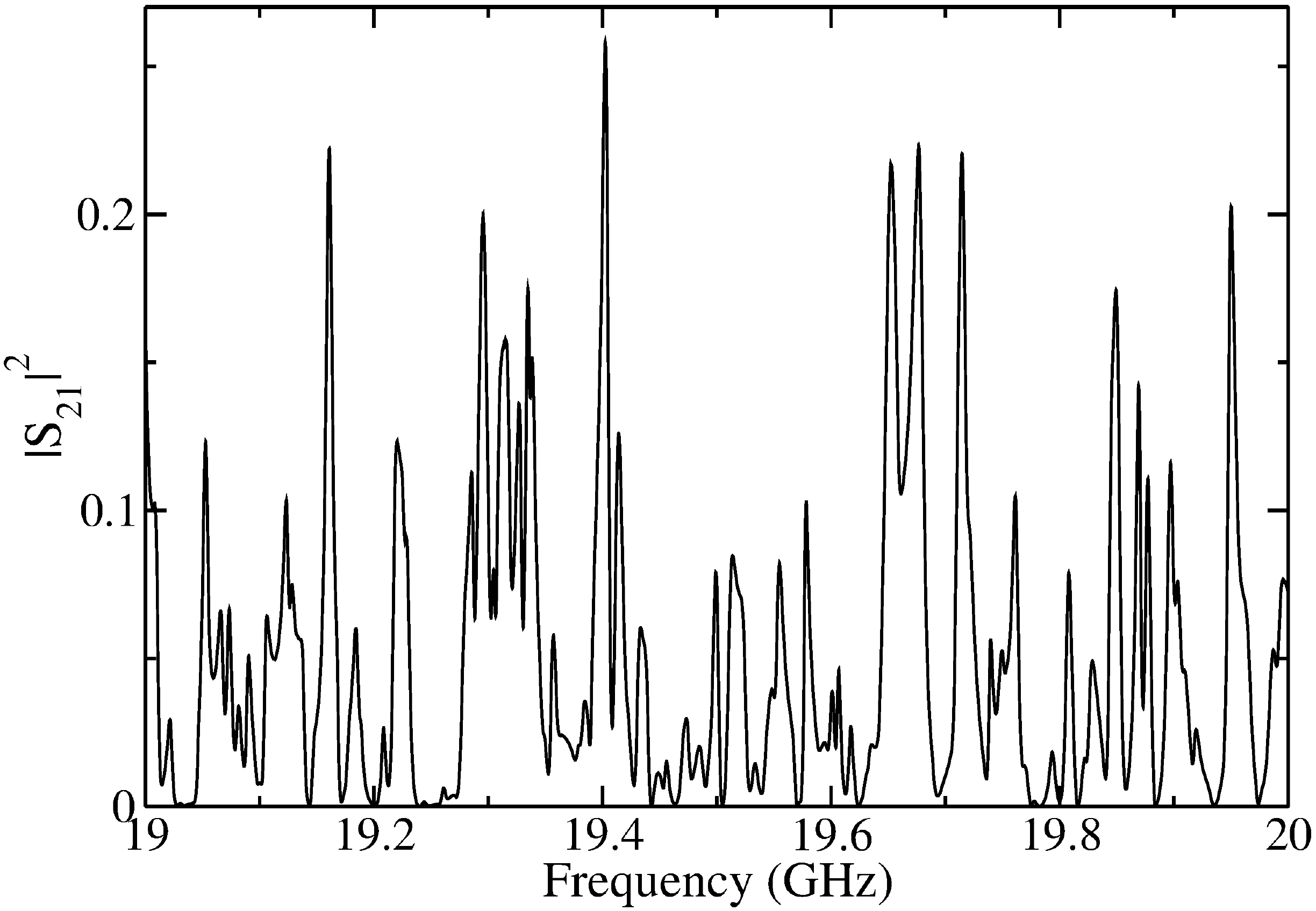}
 	\begin{tikzpicture}
        \node[anchor=south west,inner sep=0] (image) at (0,0) {\includegraphics[width=0.78\linewidth]{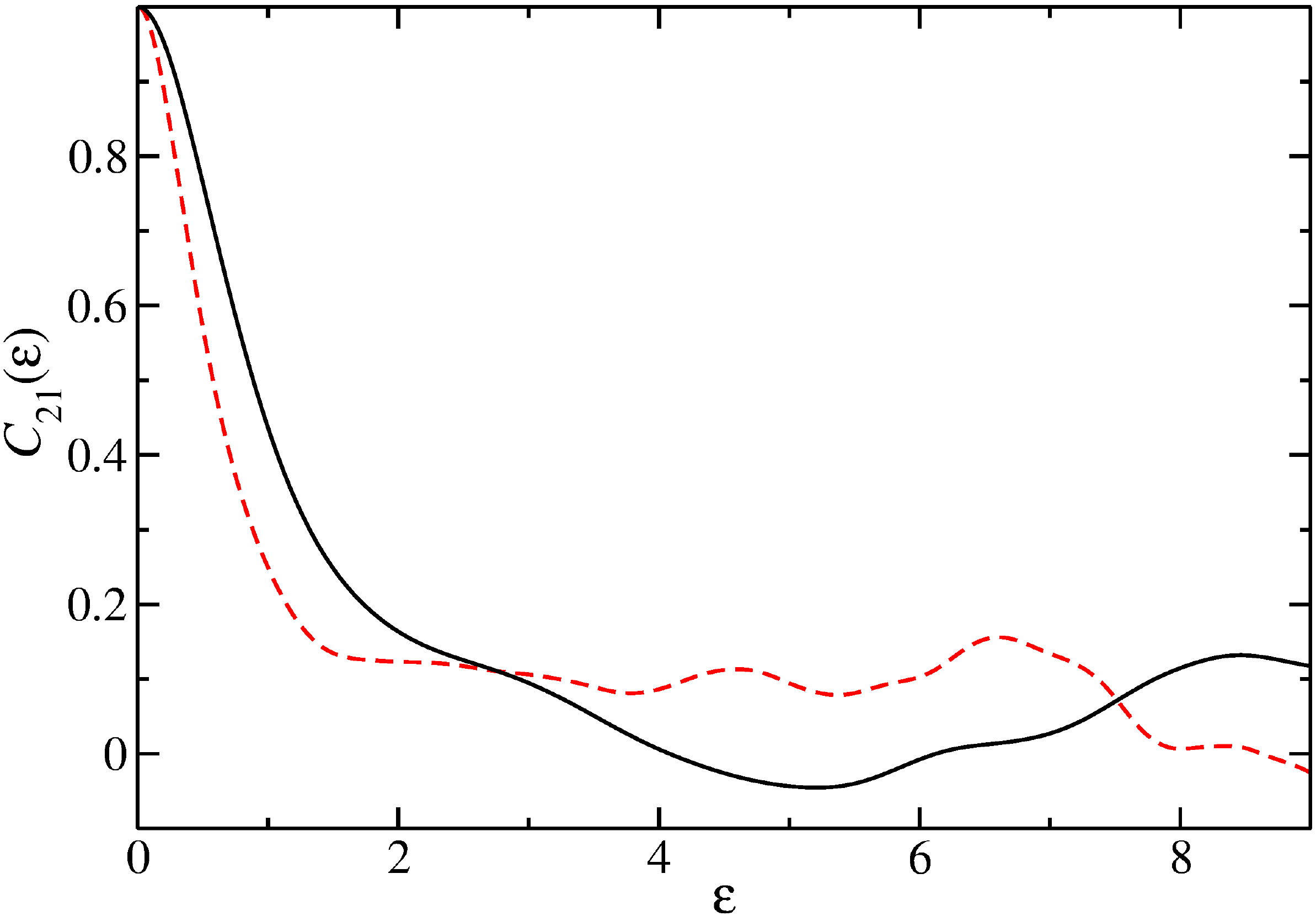}};
        \begin{scope}[x={(image.south east)},y={(image.north west)}]
            \node[anchor=south west,inner sep=0] (image) at (0.60,0.69) {\includegraphics[width=0.275\linewidth]{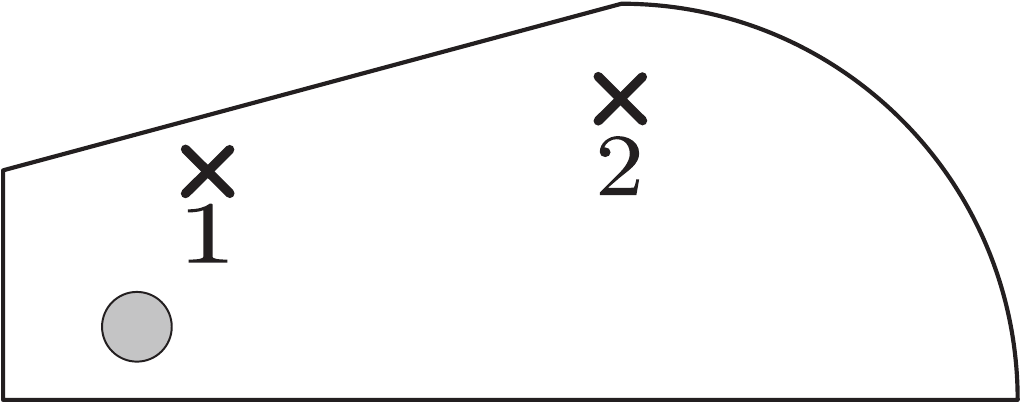}};
        \end{scope}
    \end{tikzpicture}
        \caption{Upper panel: transmission spectrum for the frequency interval $19-20$~GHz, where the resonances obviously overlap. Lower panel: cross-section autocorrelation function $\mathcal{C}_{21}(\varepsilon)$ for the same frequency interval (solid black line) and the interval of $13-14$~GHz (dashed red line). Here, $\epsilon$ gives the excitation frequency increment in units of the mean resonance spacing. The secular variation of the latter was negligible in a 1~GHz window. The broadening of the correlation width with increase in frequency is palpable. Inset: the tilted stadium billiard (schematic). The crosses mark the positions of the antennas, $1$ and $2$. A movable scatterer (gray) was utilized to obtain independent realizations.}
\label{fig1}
\end{figure}

To realize a chaotic scattering system, a flat microwave resonator with the shape of a tilted-stadium billiard~\cite{Dietz2009a,Dietz2010,Dietz2010a} was chosen, which is shown schematically in the inset of Fig.~\ref{fig1}. 
The modes in the resonator were coupled to the exterior via two antennas denoted by $1$ and $2$, that were attached to it at the positions marked by crosses. For the determination of the scattering ($S$)-matrix elements $S_{21}$ describing the scattering process from antenna $1$ to antenna $2$, a vector network analyzer coupled microwave power into the resonator via antenna 1 and determined the relative phase and amplitude of the transmitted signal at antenna 2. The frequency range of $1-25$~GHz was chosen such that only the component of the electric field strength perpendicular to the top and the bottom plate was excited. Then the Helmholtz equation is scalar and mathematically equivalent to the Schr\"odinger equation of the corresponding quantum billiard~\cite{Stoeckmann1990,Sridhar1991,Graf1992}. Therefore, such resonators are called microwave billiards. The dynamics of the corresponding classical billiard is chaotic~\cite{Primack1994}. Thus, according to the Bohigas-Giannoni-Schmit conjecture~\cite{Bohigas1984}, the fluctuation properties of the eigenvalues of the associated quantum billiard, i.e., of the resonance frequencies in the microwave billiard, are described by random matrices from the Gaussian orthogonal ensemble (GOE)~\cite{Mehta1990}. We confirmed this by analyzing the resonance spectra in the regions of isolated and overlapping resonances~\cite{Dietz2010}. An ensemble of several chaotic systems was obtained by introducing a scatterer (gray disk in the inset of Fig.~\ref{fig1}) into the microwave billiard and moving it to 8 different positions~\cite{Mendez-Sanchez2003}. We also investigated fluctuation properties in chaotic scattering systems with partially broken time-reversal symmetry. To induce \T violation we inserted a ferrite into the microwave billiard and magnetized it with an external magnetic field.

In the upper panel of Fig.~\ref{fig1} a measured transmission spectrum, i.e., the squared modulus of the $S$-matrix element from antenna 1 to antenna 2, is plotted in the frequency range $19-20$~GHz. The ratios of the widths and the spacings of the resonances depend weakly on the excitation frequency. At low frequencies the resonances were isolated, that is, their mean width was small as compared to their mean spacing. However, on increasing it, the resonances started to overlap, as is clearly visible in Fig.~\ref{fig1}. The lower panel of Fig.~\ref{fig1} shows two examples for the correlation function of the squared modulus of the $S$-matrix element $S_{21}$, which corresponds to the cross section for the transition from channel 1 to channel 2 in nuclear physics, $\sigma_{21}=\vert S_{21}\vert^2$. It is given by~\cite{Ericson1963,Bonetti1983}, 
\begin {equation}
\mathcal{C}_{21}(\varepsilon) = \langle\lvert S_{21}(E)\rvert^2\lvert S_{21}(E+
\varepsilon) \rvert^2 \rangle - \langle \lvert S_{21} \rvert^2 \rangle^2,
\label{eq:cross}
\end{equation}
where $\langle\ldots \rangle$ denotes averaging over $E$. In nuclear physics $E$ and $E+\varepsilon$ denote energies, whereas in the microwave experiments they correspond to the rescaled frequencies $f$, obtained by unfolding them to mean resonance spacing unity with the help of Weyl's formula for the smooth part of the resonance density $\langle\rho\rangle$ of microwave billiards~\cite{Weyl1912} which increases linearly with the frequency $f$, $\langle\rho\rangle\simeq A\pi f/(2c^2)$ with $c$ the velocity of light and $A$ the area of the billiard. In the limit of a large number of open channels the shape of the cross-section correlation function $\mathcal{C}_{21}(\varepsilon)$ is well decribed by a Lorentzian, $\mathcal{C}_{21}(\varepsilon)=\mathcal{C}_{21}(0)/\left(1+\left(\varepsilon/\Gamma_{\rm corr}\right)^2\right)$.

The averages of the resonance widths and spacings were observed to be approximately constant in 1~GHz frequency intervals. Consequently, the correlation function was evaluated in a frequency range of $5-25$~GHz in 1~GHz windows to ensure that the secular variation of the resonance parameters is negligible, yielding $10^4$ data points each. As illustrated in Fig.~\ref{fig1}, the width of the cross-section function increases with the frequency, that is, it is narrower in the frequency range $13-14$~GHz (dashed red line) than for $19-20$~GHz (solid black line).  

The number of maxima $N^{\max}$ in the transmission spectra was also counted in $1$~{\rm GHz} windows. One thereby obtains the density of maxima, $\rho^{\max}_{\varepsilon}$, on simply dividing $N^{\max}$ by the corresponding interval length of the rescaled frequencies. Furthermore, each of the $1$~{\rm GHz} windows was subdivided into 10 intervals and in each of them the width of the cross-section autocorrelation function at half maximum, i.e., the value of $\varepsilon$ with $\mathcal{C}_{21}(\varepsilon)=0.5\, \mathcal{C}_{21} (0)$, was determined in units of the mean resonance spacing. To further increase the data sets, transmission spectra were measured for 8 positions of the movable scatterer.
\begin{figure}[h]
    \begin{tikzpicture}
        \node[anchor=south west,inner sep=0] (image) at (0,0) {\includegraphics[width=0.8\linewidth]{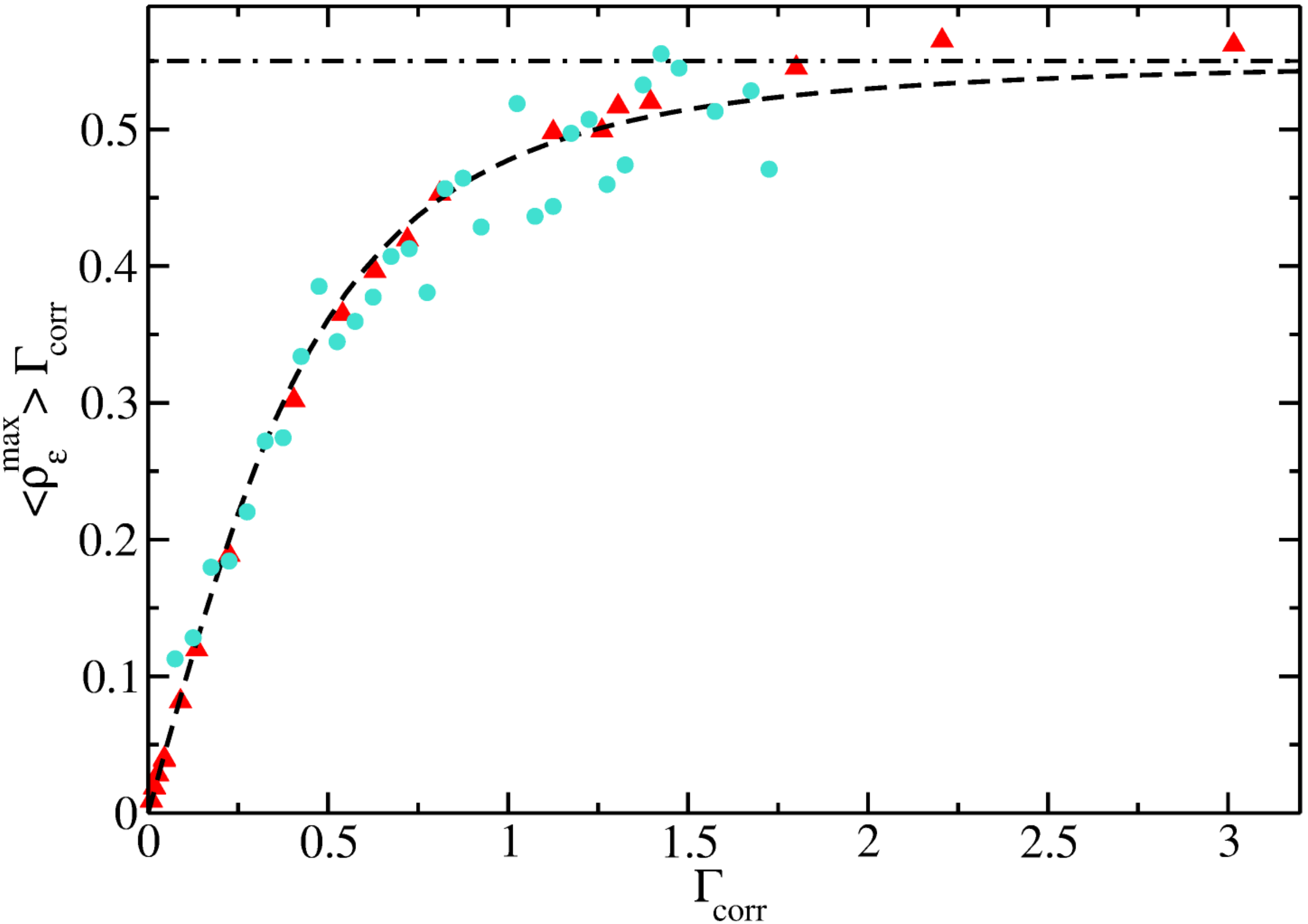}};
        \begin{scope}[x={(image.south east)},y={(image.north west)}]
                 \node[anchor=south west,inner sep=0] (image) at (0.60,0.20) {\includegraphics[width=0.275\linewidth]{Tilted_Ohne_Ferrit_kl.pdf}};
        \end{scope}
    \end{tikzpicture}
    \caption{Product of the mean density of maxima and the mean correlation width in the transmission spectra of the tilted-stadium billiard shown in the inset (same as in Fig.~\ref{fig1}) times the mean correlation width evaluated in 1~GHz windows in the $\mathcal{T}$ preserved case (turquoise dots). Statistical errors arise due to the variability of the results within the 8 realizations, i.e., positions of the scatterers, and are at most of the size of the symbols. Therefore, we omitted error bars. The red triangles, obtained from RMT simulations with an ensemble of scattering matrices of the form Eq.~(\ref{eqn:Sab}) for the GOE case describe the experimental results well. The value of $0.5 b_\infty=\sqrt{3}/\pi\approx 0.55$ (dash-dotted line) approached above $\Gamma_{\rm corr}\approx 1.5$ is in good agreement with the result obtained in Ref.~\cite{Brink1963} in the context of nuclear physics. The dashed curve illustrates the analytical expression~\cite{Barbosa2013} given in Eq.~(\ref{eqn:hussein1}) with $a_0=1$ and $a_1=1/3$.} 
\label{fig2}
\end{figure}

Figure~\ref{fig2} presents the thus obtained experimental results for the product of the mean density of maxima and the mean correlation width (turquoise dots). The largest value of the latter achieved in the experiments was $\Gamma_{\rm corr}\simeq 1.7$. We demonstrated in Ref.~\cite{Dietz2010a} that in the corresponding frequency range the $S$-matrix correlation functions and the distributions of the $S$-matrix elements are well described by the predictions for the Ericson region. Similarly, the product of the  mean density of maxima and the mean correlation width approaches the value predicted in that limit, $\langle \rho^{\max}_{\varepsilon} \rangle\Gamma_{\rm corr}\simeq 0.5\, b_\infty=\sqrt{3}/\pi$, shown as dash-dotted line in Fig.~\ref{fig2}. Note, that $\langle \rho^{\max}_{\varepsilon}\rangle\Gamma_{\rm corr}$ does not depend on the frequency scale. 

\begin{figure}[h]
    \begin{tikzpicture}
        \node[anchor=south west,inner sep=0] (image) at (0,0) {\includegraphics[width=0.8\linewidth]{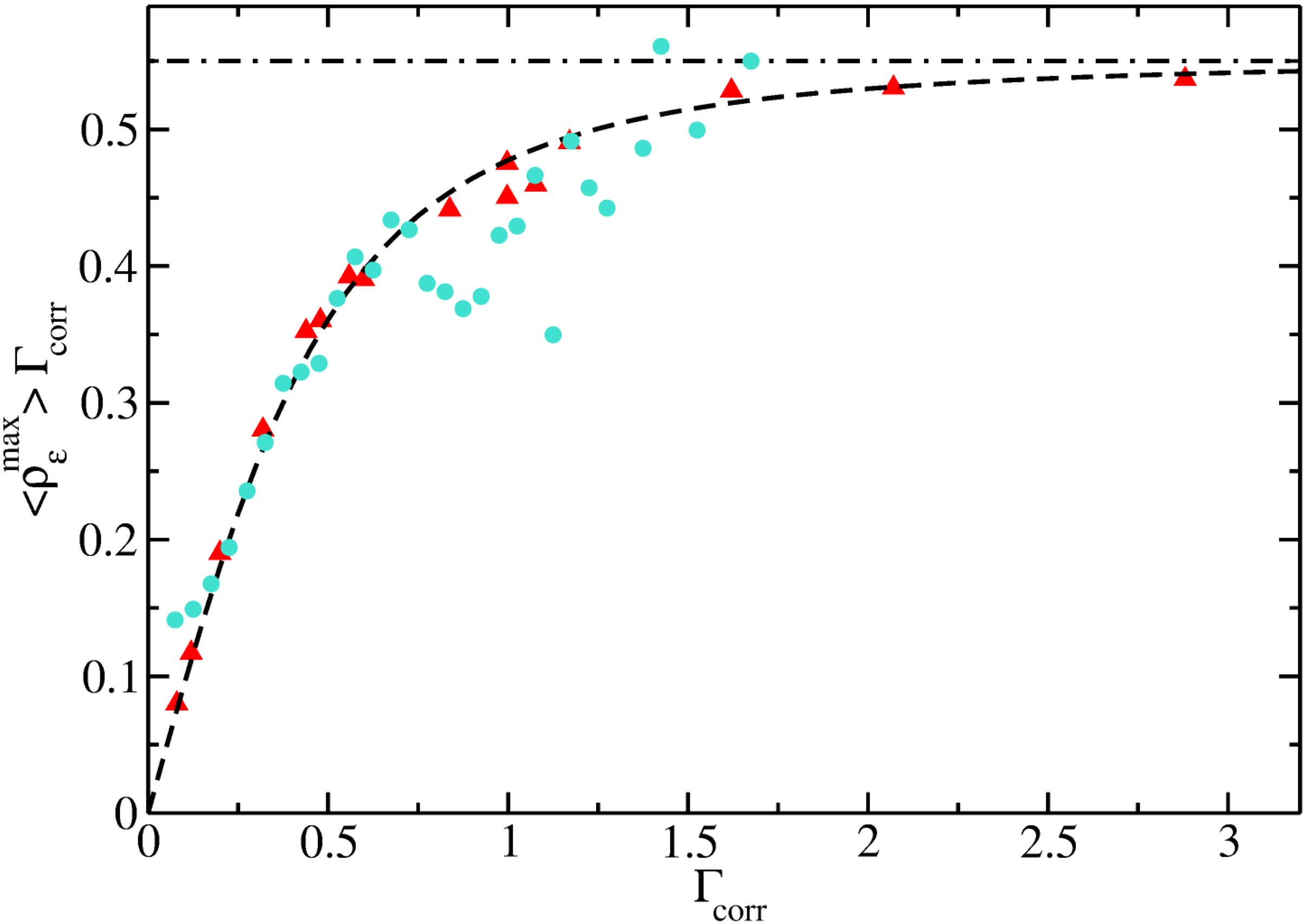}};
        \begin{scope}[x={(image.south east)},y={(image.north west)}]
                 \node[anchor=south west,inner sep=0] (image) at (0.60,0.20) {\includegraphics[width=0.275\linewidth]{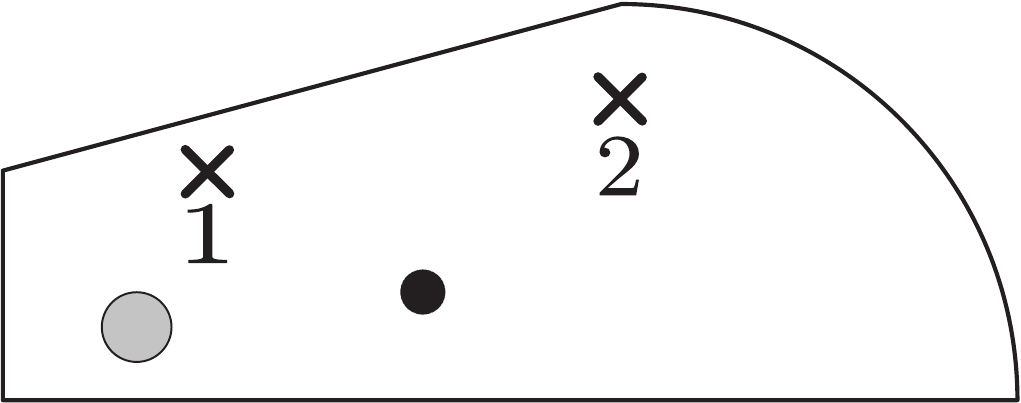}};
        \end{scope}
    \end{tikzpicture}
        \caption{Same as Fig.~\ref{fig2} for the cases of \T violation in the experiment (turquoise dots) and the GUE in the RMT simulations (red triangles). Inset: same as in Fig.~\ref{fig1}. In addition a magnetized ferrite (black disk) was inserted to enforce $\mathcal{T}$ violation.}
\label{fig3}
\end{figure}

In order to obtain an analytical expression for the dependence of the product $\langle\rho^{max}_{\varepsilon}\rangle\Gamma_{\rm corr}$ on the correlation width $\Gamma_{\rm corr}$ we used the ansatz  
\begin{equation}
\langle \rho^{\max}_{\varepsilon}\rangle\Gamma_{\rm corr}\approx \frac{\sqrt{3}}{\pi}a_0\frac{\Gamma_{\rm corr}}{\sqrt{\Gamma_{\rm corr}^2+a_1}}, 
\label{eqn:hussein1}
\end{equation}
with $a_0$ and $a_1$ as fit parameters. The comparison of our experimental data with the analytical expression Eq.~(\ref{eqn:hussein1}) revealed a very good agreement for $a_0=1$ and $a_1=1/3$. Hence, like in the case of the experimental data, the analytical expression saturates at the value predicted for the Ericson region. It is shown as dashed line in Fig.~\ref{fig2} and also in Fig.~\ref{fig3}, where the product $\langle \rho^{\max}_{\varepsilon}\rangle\Gamma_{\rm corr}$ is plotted for the case with \T violation, induced by inserting a ferrite into the microwave billiard (black disk in Fig.~\ref{fig3}) and magnetizing it with an external magnetic field~\cite{Dietz2007a,Dietz2009a}. The agreement of the analytical expression Eq.~(\ref{eqn:hussein1}) and the experimental result for $\langle \rho^{\max}_{\varepsilon}\rangle\Gamma_{\rm corr}$ again is very good for $a_0=1$ and $a_1=1/3$. 
No significant difference is noticeable between the cases of \T preservation and \T violation, either in the saturation value or in the general trend of the data points.

\begin{figure}[h]
\includegraphics[width=\linewidth]{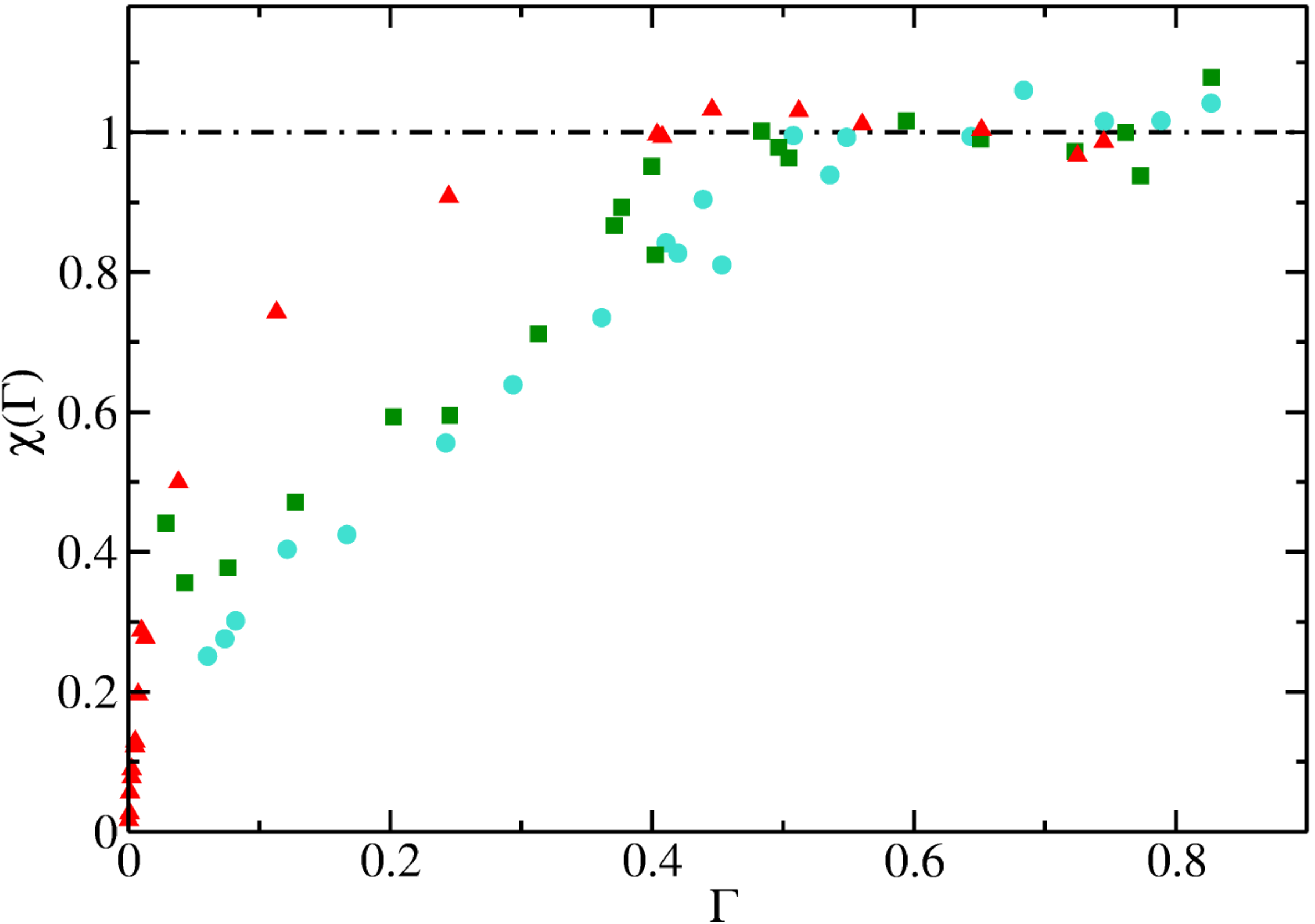}
\caption{(Color on line) Ratio $\chi(\Gamma)=\langle\rho^{\rm max}_{\varepsilon}\rangle\Gamma_{\rm corr}/(\langle\rho^{\rm QD}_{\varepsilon}\rangle\gamma)$ of the experimental result $\langle\rho^{\rm max}_{\varepsilon}\rangle\Gamma_{\rm corr}$ for the cases of \T preservation and \T violation (turquoise dots and green squares, respectively) and the analytical result for a quantum dot, see Eq.~(\ref{eqn:hussein1ex}). Furthermore, the ratio $\chi(\Gamma)=\langle\rho^{\max}_{X}\rangle X_C/(\langle\rho^{\rm QD}_X\rangle X_C)$ of the RMT simulation for the parameter dependent system $\langle\rho^{\max}_{X}\rangle X_C$ and the analytical result Eq.~(\ref{eqn:hussein2ex}) for a quantum dot is plotted (red triangles). Here, $\Gamma$ denotes the average of the transmission coefficients associated with the two antennas 1 and 2, $\Gamma=(T_1+T_2)/2$, in the case of the microwave billiard and the tunneling probability in that of a quantum dot.}
\label{fig8}
\end{figure}

In Ref.~\cite{Hussein2014} the conductance fluctuations where investigated in quantum dots with chaotic behaviour. There, an analytical expression was derived for the product of the mean density of maxima $\langle\rho^{\rm QD}_{\varepsilon}\rangle$ and the width $\gamma$ of the generalized conductance correlation function which, similarly to $C_{21}(\epsilon)$ in Eq.~(\ref{eq:cross}), takes the shape of a Lorentzian in the Ericson region,
\begin{equation}
\langle\rho^{\rm QD}_{\varepsilon}\rangle\gamma =\frac{\sqrt{3}}{\pi}\sqrt{\frac{9\Gamma^2-18\Gamma+10}{5\Gamma^2-a_1\Gamma+6}}.
\label{eqn:hussein1ex}
\end{equation}
Here, $\Gamma$ denotes the tunneling probability, which corresponds to the transmission coefficients $T_a=1-\vert\langle S_{aa}\rangle\vert^2$ associated with antennas $a=1,2$ in our microwave experiments, $\Gamma =\frac{T_1+T_2}{2}$, where $T_1\simeq T_2$. In order to verify Eq.~(\ref{eqn:hussein1ex}) we plotted in Fig.~\ref{fig8} the ratios $\chi(\Gamma )$ of the experimental results for $\langle \rho^{\max}_{\varepsilon}\rangle\Gamma_{\rm corr}$ for the cases of preserved \T (turquoise dots) and of \T violation (red triangles) and the prediction Eq.~(\ref{eqn:hussein1ex}), $\chi(\Gamma)=\langle\rho^{\rm max}_{\varepsilon}\rangle\Gamma_{\rm corr}/(\langle\rho^{\rm QD}_{\varepsilon}\rangle\gamma)$.
For $\Gamma\gtrsim 0.4$ the ratio takes the value $\chi(\Gamma)\simeq 1$, i.e., there the agreement between the experimental result and the prediction Eq.~(\ref{eqn:hussein1ex}) is very good. For small values of $\Gamma$ deviations are clearly visible and, actually, expected because for $\Gamma\to 0$ the product $\langle\rho^{\max}_{\varepsilon}\rangle\Gamma_{\rm corr}$ vanishes, see Figs.~\ref{fig2} and~\ref{fig3}, whereas the square-root function in Eq.~(\ref{eqn:hussein1ex}) takes a non-zero value for $\Gamma\to 0$~\cite{Ramos2011}. 
\section{\label{RMT} Fluctuations in a random $S$-matrix model for chaotic scattering processes} 
To corroborate our experimental findings we performed RMT simulations similar to those presented in Refs.~\cite{Dietz2009a,Dietz2010}. For this we used the $S$-matrix formalism developed in~\cite{Mahaux1969} in the context of compound-nucleus reaction theory, which, actually, also has been used in~\cite{Ramos2011,Barbosa2013} for the RMT simulations. The associated unitary $S$ matrix has the general form
\begin{equation}
        S(f) = \II - 2\pi i W \big(f\II-H + i\pi W^T W\big)^{-1} W^T\, .
        \label{eqn:Sab}
\end{equation}
The matrix elements of $W$ describe the coupling of the antennas to the resonator modes~\cite{Dietz2007a} and comprise the fictitious channels that account for the Ohmic absorption in the cavity~\cite{RSchaefer2003}. They were chosen as real Gaussian distributed random numbers with zero mean. As stated above, the fluctuation properties of the resonance frequencies of the resonator are well described by random GOE matrices in the \T-preserving case. Thus we inserted in Eq.~(\ref{eqn:Sab}) for the matrix $H$, which actually stands for the Hamiltonian of the closed resonator, a real and symmetric random matrix of dimension $N=200$ from the GOE~\cite{Mehta1990}. In order to study a chaotic scattering system with induced partial \T violation we proceeded as in Refs.~\cite{Dietz2009a,Dietz2010} and inserted for $H$ in Eq.~(\ref{eqn:Sab}) a complex random matrix of the form $H=H^S+i\pi\xi/\sqrt{N}H^A$, with $H^S$ a member from the GOE, $H^A$ a real, antisymmetric matrix with Gaussian-distributed entries and $\xi$ the \T-breaking parameter. The number of open channels $M$, that is, the dimension of $S(f)$ was chosen equal to 32, i.e., we introduced 30 fictitious ones. Furthermore, since in the experiments the average $S$ matrix was diagonal, we required that this is also the case for the matrix $WW^T$~\cite{Mahaux1969,Verbaarschot1985}. Its entries are solely determined by the transmission coefficients $T_c$, with $c=1,\cdots,M$. For the RMT simulations we used the same values of the transmission coefficients as in the experiments within the 1~GHz frequency intervals, see Ref.~\cite{Dietz2010}. 

The RMT results plotted as red triangles in Figs.~\ref{fig2} and~\ref{fig3} were obtained with an ensemble of $1000$ random scattering matrices of the form Eq.~(\ref{eqn:Sab}). The data presented in Fig.~\ref{fig3} were generated by choosing $\xi$ as in the measurements~\cite{Dietz2009a,Dietz2010}, where it varied with the excitation frequency. The RMT results for $\xi=0$ and $\xi\ne 0$ are indistinguishable. Thus, the value of the product $\langle\rho^{\rm max}_{\varepsilon}\rangle\Gamma_{\rm corr}$ does not depend on whether \T is preserved or not, in agreement with the experimental results. It saturates at $0.5\, b_{\infty}=\sqrt{3}/\pi$, i.e., at the value obtained in the Ericson region in the context of nuclear physics for systems with a large number of reaction channels or open channels~\cite{Dallimore1965}. Furthermore, the RMT results are again well described by the formula Eq.~(\ref{eqn:hussein1}) with $a_0=1$ and $a_1=1/3$. 

\section{\label{Graph} Fluctuations in open quantum graphs}
\begin{figure}[h]
        \begin{tikzpicture}
        \node[anchor=south west,inner sep=0] (image) at (0,0) {\includegraphics[width=0.8\linewidth]{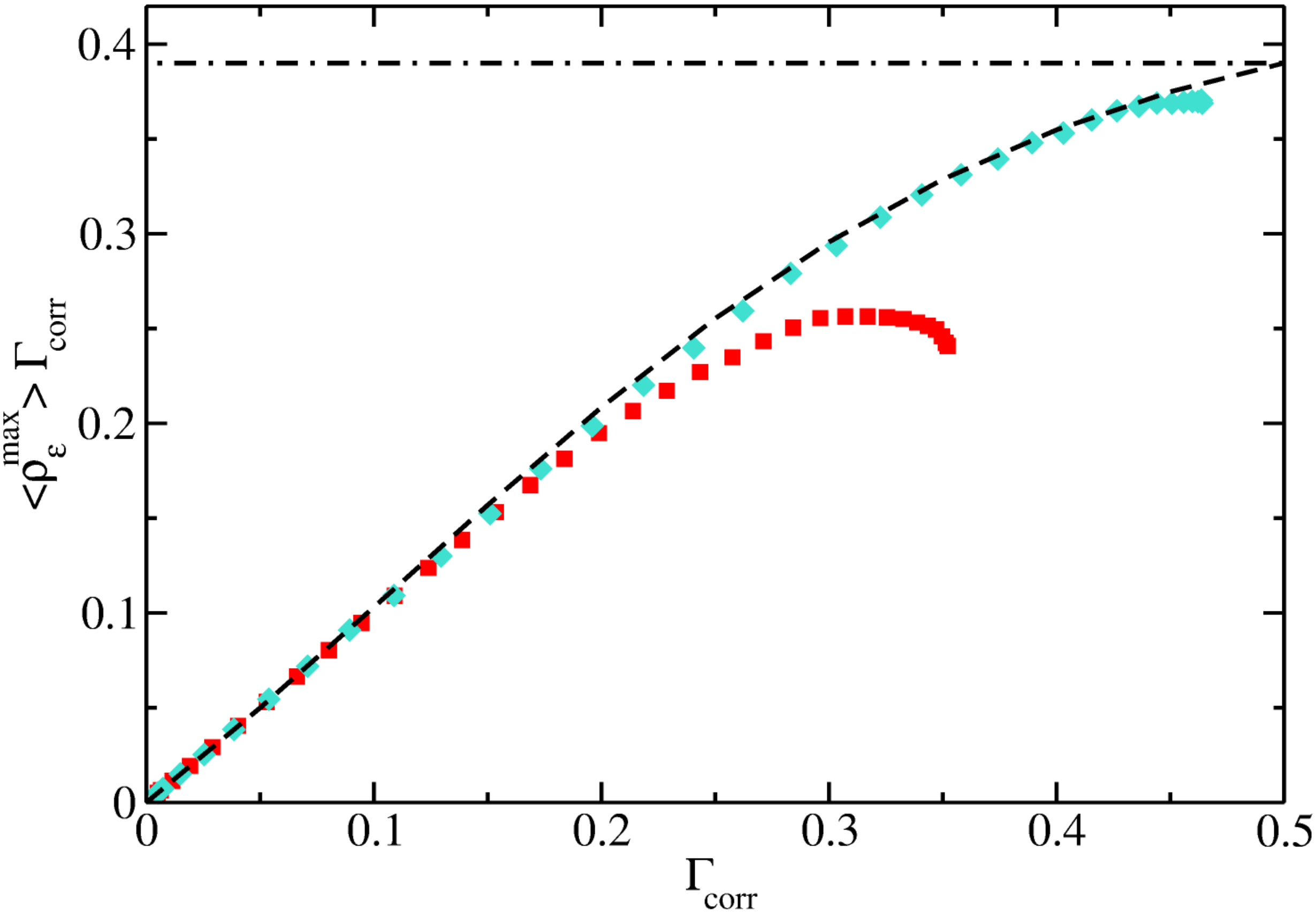}};
        \begin{scope}[x={(image.south east)},y={(image.north west)}]
            \node[anchor=south west,inner sep=0] (image) at (0.60,0.20) {\includegraphics[width=0.275\linewidth]{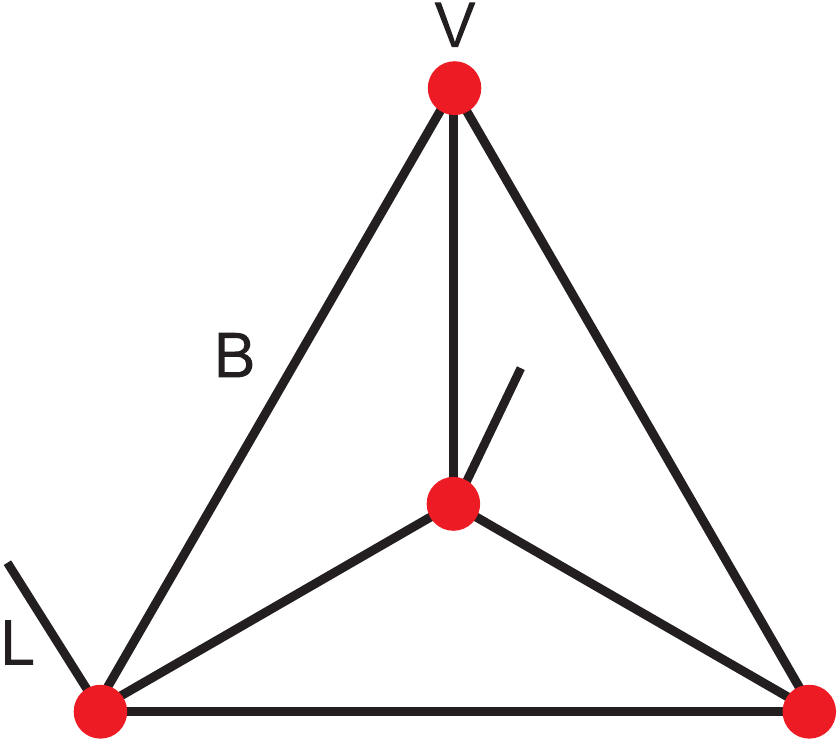}};
        \end{scope}
    \end{tikzpicture}
        \caption{Product of the mean density of maxima and the mean correlation width for a chaotic tetrahedral graph shown schematically in the inset, which consists of 6 bonds B, 4 vertices V and has two leads L attached, corresponding to two open channels, obtained by varying the coupling $w$ between the leads and the vertices. Shown are the results for the case of \T preservation (turquoise dots) and for complete \T violation (red squares), where for all bonds $A_{ij}=0.5,\, i<j$ was used. The data are well described by Eq.~(\ref{eqn:hussein11}) with $d_0=1,\, d_1=1/3$ and $d_2=1/6$ (dashed line). For large $\Gamma_{\rm corr}$ the product approaches the value $0.5\, b_2=0.39$ (dash-dotted line) predicted for nuclear reactions with 2 independent channels~\cite{Brink1963}. Note that the graph shown in the inset is merely representational since the lengths do not reflect those used in the actual simulations.}
\label{fig4}
\end{figure}

As a second test system we analyzed the $S$ matrix of open quantum graphs, networks of one-dimensional wires (bonds) joined at vertices~\cite{Londergan1999} and opened by attaching leads that extend to infinity at some of the vertices. An example is shown in the inset of Fig.~\ref{fig4}, where a so-called tetrahedral graph is presented, which consists of 6 bonds B, 4 vertices V and has a lead L attached to the lower left vertix and to the center one, respectively. 
Quantum graphs can be handled mathematically and at the time serve as a simple model for chaotic scattering. They, in fact, have the particular property that the semiclassical approximation is exact~\cite{Keating1991} and, moreover, the $S$-matrix correlation functions could be calculated explicitely and shown to agree with the RMT result~\cite{Pluhar2013,Pluhar2014,Verbaarschot1985,Fyodorov2005}. The mathematical model for the $S$ matrix of open quantum graphs has been formulated in Refs.~\cite{Kottos1999,Texier2001}. The component $\psi_{ij}$ of the graph wavefunction $\psi$ for the mode propagation in the bond from vertex $i$ to vertex $j$ satisfies the one-dimensional Schr\"{o}dinger equation
\begin{equation}
\bigg( -\mathrm{i} \frac{\operatorname{d}}{\mathrm{dx}} - A_{ij} \bigg)^2 \psi_{ij} (x) = k^2 \psi_{ij} (x),
\end{equation}
where $A_{ij}=-A_{ji}$ is a magnetic vector potential with a nonvanishing real part introduced on the bonds to induce \T violation. We set it to zero for the study of chaotic scattering in \T-preserved systems, and chose it purely real with $\lvert A_{ij} \rvert$ being the same on all bonds to induce \T violation, otherwise. Furthermore, we required that the wave functions are continuous and satisfy Neumann boundary conditions at the vertices in order to ensure current conservation. On attaching a lead to $M$ of the in total $V$ vertices of a graph, the $S$ matrix can be expressed as
\begin{equation}
S(f) = \II-2\pi\mathrm{i}W\big(h(f) + i\pi W^{T}W\big)^{-1}W^T,
\label{SGraph}
\end{equation}
with $W$ being the $M\times V$ leads-vertices coupling matrix introduced in Ref.~\cite{Texier2001} for a tunable coupling~\cite{Exner1997}, i.e., its matrix elements depend on a parameter $w$, $0 < W_{i_L,j} = \delta_{i_L,j}w/\sqrt\pi\leq 1$, where $i_L$ denotes the indices of the vertices with leads. The term $h(f)$ is the Hamiltonian of the closed graph with the matrix elements 
\begin{equation}
h_{ij} (f) =\begin{cases}
    -\Sigma_{m\ne i}\, C_{im} \cot\left(\frac{2\pi f}{c}L_{im}\right), & \text{$i = j$}\\
    C_{ij}\, \mathrm{e}^{-\mathrm{i} A_{ij} L_{ij}}\,\sin^{-1}\left(\frac{2\pi f}{c}L_{ij}\right), & \text{$i \ne j$}
  \end{cases}.
\end{equation}
Here, $C$ is the connectivity matrix with entries $C_{ij}=1$ if the vertices $i$ and $j$ are connected, and zero otherwise. Furthermore, $L$ is the length matrix of the graph, where $L_{ij}$ gives the length of the bond joining the vertices $i$ and $j$. Note the similarity of the form of the $S$ matrix of an open graph, Eq.~(\ref{SGraph}), with that of a compound-nucleus scattering process given in Eq.~(\ref{eqn:Sab}), the only difference being their dependence on the frequency through the inverse operator. In the former case it coincides with that of the graph Hamiltonian $h(f)$, whereas in the latter case it depends linearly on the frequency, while the Hamiltonian of the closed system is modelled by a frequency-independent random matrix. 

\begin{figure}[h]
       \begin{tikzpicture}
        \node[anchor=south west,inner sep=0] (image) at (0,0) {\includegraphics[width=0.8\linewidth]{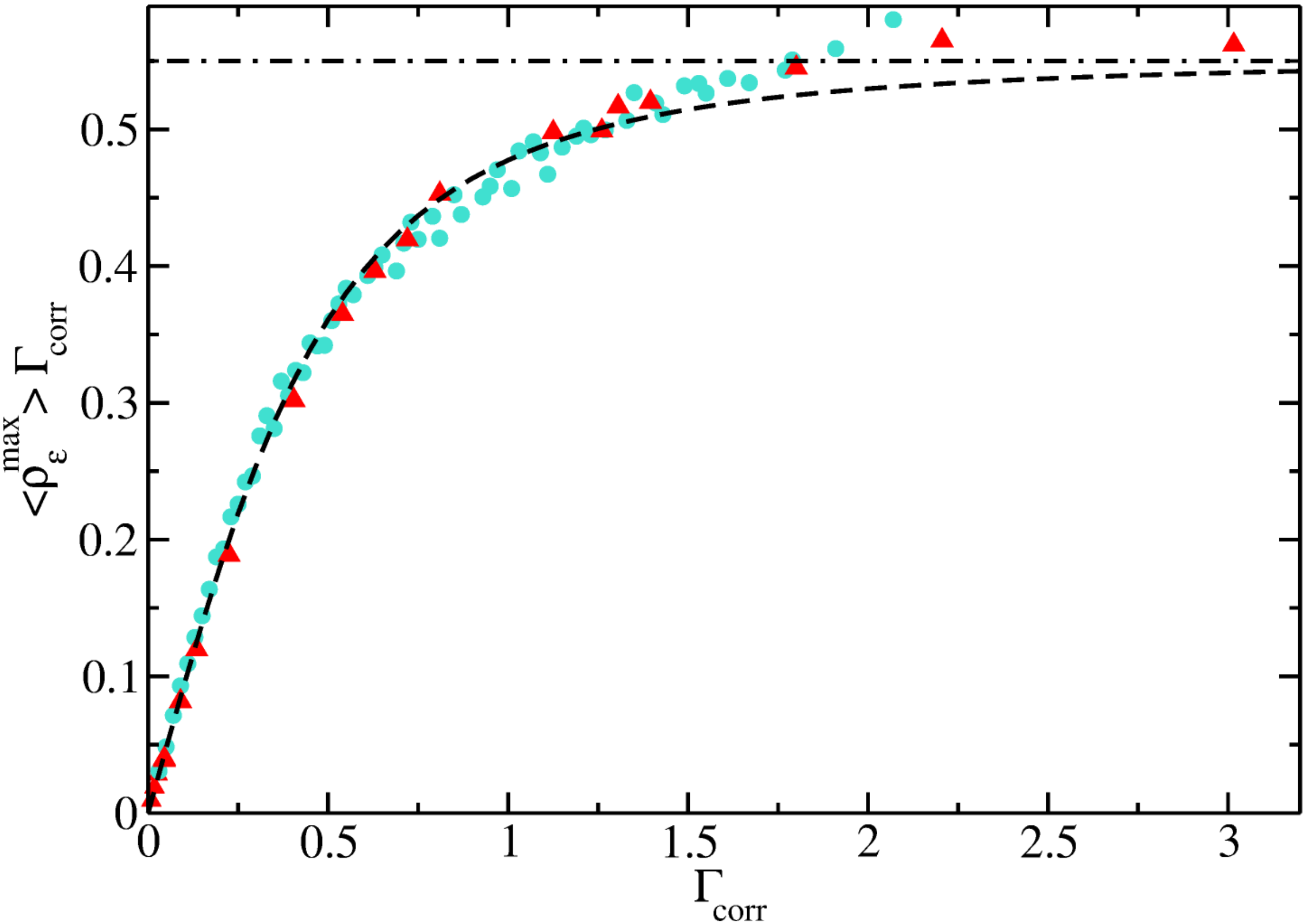}};
        \begin{scope}[x={(image.south east)},y={(image.north west)}]
            \node[anchor=south west,inner sep=0] (image) at (0.60,0.20) {\includegraphics[width=0.275\linewidth]{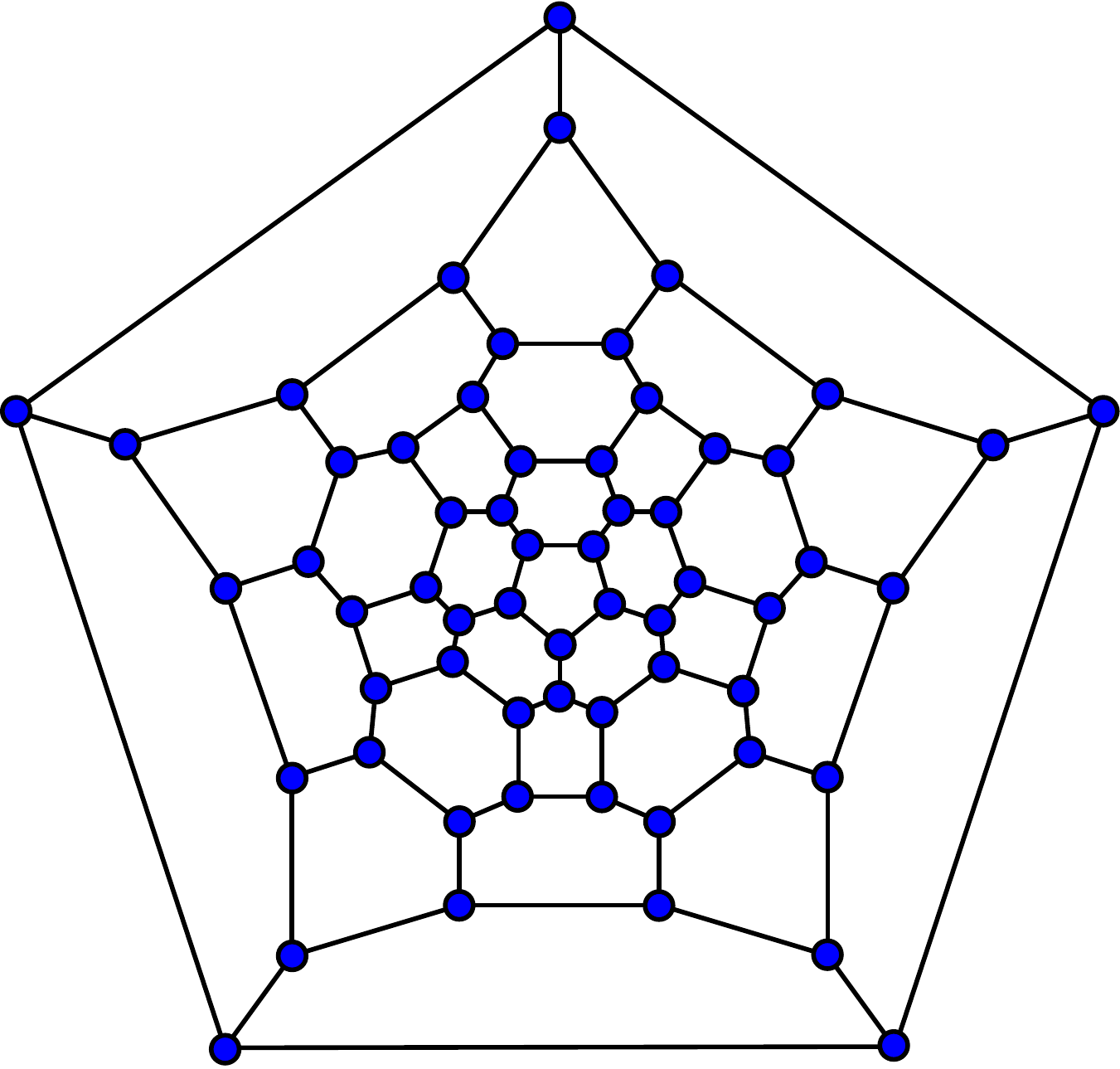}};
        \end{scope}
    \end{tikzpicture}
        \caption{Product of the mean density of maxima and the mean correlation width obtained from simulations performed for a chaotic graph with $V=60$, shown schematically in the inset, and a varying number of open channels ($2-15$) and couplings $w$ between the leads and the vertices are illustrated as turquoise dots. The red triangles are the same as those exhibited in Fig.~\ref{fig2}. A saturation at $0.5 \, b_{\infty} = 0.55$ is observed, in agreement with the value cited in Ref.~\cite{Dallimore1965}. Furthermore, the data are well described by Eq.~(\ref{eqn:hussein1}) with $a_0=1,\, a_1=1/3$. Note that the equality of the side-lengths of the graph shown in the inset is solely for representational convenience.}
\label{fig5}
\end{figure}

We considered in our numerical simulations tetrahedral graphs consisting of 4 vertices and graphs that comprised 60 vertices, where in both cases the connectivity was 3 (see insets of Figs.~\ref{fig4} and~\ref{fig5}). The lengths of the bonds were confirmed to be rationally independent by choosing values which were the square roots of prime numbers. In addition, the chaoticity of the scattering dynamics was independently verified by the fluctuation properties of the eigenvalues of the closed graphs. The $S$-matrix elements were computed over a large frequency range and the frequencies were again rescaled to mean resonance spacing one using Weyl's formula for the mean resonance density $\langle\rho\rangle$ in quantum graphs, which is frequency independent, $\langle\rho\rangle\simeq\mathcal {L}/(2\pi)$, with $\mathcal{L}$ the total length of the graph. Consequently the mean resonance spacing is the same for the whole spectrum, in stark contrast to the situation in quantum (microwave) billiards. The same holds for the correlation width. Accordingly an ensemble of statistically indendent data sets was obtained by simply dividing the spectrum into frequency windows containing a sufficiently large number of resonances, determining in each of them the number of maxima and the mean width of the cross-section correlation function and then calculating their arithmetic averages. 

In the case of the tetrahedral graph shown schematically in the inset of Fig.~\ref{fig4} we attached to two vertices a lead and varied the size of $\Gamma_{\rm corr}$ by altering the parameter $w$ of the entries of the $W$ matrix, which physically corresponds to changing the strength of the coupling between the leads and the vertices or that of the graph to its environment.  
In Fig.~\ref{fig4} the quantity $\langle \rho^{\max}_{\varepsilon}\rangle \Gamma_{\rm corr}$ is plotted against $\Gamma_{\rm corr}$ for the \T-preserving (turquoise diamonds) and the \T-violated (red squares) cases. Especially for the \T-invariant graph its saturation at $0.5 \, b_2 \approx 0.39$ (dash-dotted line) is clearly visible. This value was predicted for 2 reaction channels in nuclear physics~\cite{Brink1963} and for $2-3$~open channels in quantum dots~\cite{Ramos2011}. Again, the product is well described by an expression similar to Eq.~(\ref{eqn:hussein1}),
\begin{equation}
\langle \rho^{\max}_{\varepsilon}\rangle\Gamma_{\rm corr}\approx 0.39d_0\frac{\Gamma_{\rm corr}}{\sqrt{\Gamma_{\rm corr}^2+d_1\Gamma_{\rm corr}+d_2}}.
\label{eqn:hussein11}
\end{equation}
The fit of Eq.~(\ref{eqn:hussein11}) to the numerical data yielded $d_0=1,\, d_1=-1/3,\, d_2=1/6$ (dashed line). Thus, the analytical expression saturates at the same value, $0.5 b_2$, as the experimental data.

In order to attain higher values of $\Gamma_{\rm corr}$, we used the fact that it increases with the number of vertices $V$~\cite{Kottos2003}. Accordingly, we considerered a graph with $V=60$ which also accorded the flexibility of increasing the number of open channels. The product $\langle \rho^{\max}_{\varepsilon} \rangle \Gamma_{\rm corr}$ was calculated for different values of the correlation width by attaching leads at random to a varying number of $2-15$~vertices and also altering the value of $w$ in the elements of the $W$ matrix.
Figure \ref{fig5}, which compares the resulting values with the RMT simulations already presented in Fig.~\ref{fig2}, reiterates the agreement with Eq.~(\ref{eqn:hussein1}) with $a_0=1,\, a_1=1/3$ shown as dashed line and, furthermore, with the results obtained in nuclear physics a few decades earlier (dash-dotted line) for $\Gamma_{\rm corr}\gtrsim 1.75$, and highlights the analogy between open channels in generic chaotic scattering systems and the reaction channels in compound-nucleus scattering processes. The results presented in this section were obtained by analyzing numerically obtained $S$-matrix elements. Currently, we build up an experimental setup for high-precision measurements of the $S$-matrix elements of quantum graphs using networks of superconducting waveguides. 

In the previous sections~\ref{Exp1},~\ref{RMT} and~\ref{Graph} we investigated the fluctuations exhibited by the cross sections when varying the excitation frequency. In the following section we study the counting-of-maxima method in a system that depends on a geometric parameter.  

\section{\label{Exp2} Fluctuations in an open parametric microwave billiard} 
\begin{figure}[h]
     \begin{tikzpicture}
        \node[anchor=south west,inner sep=0] (image) at (0,0) {\includegraphics[width=0.8\linewidth]{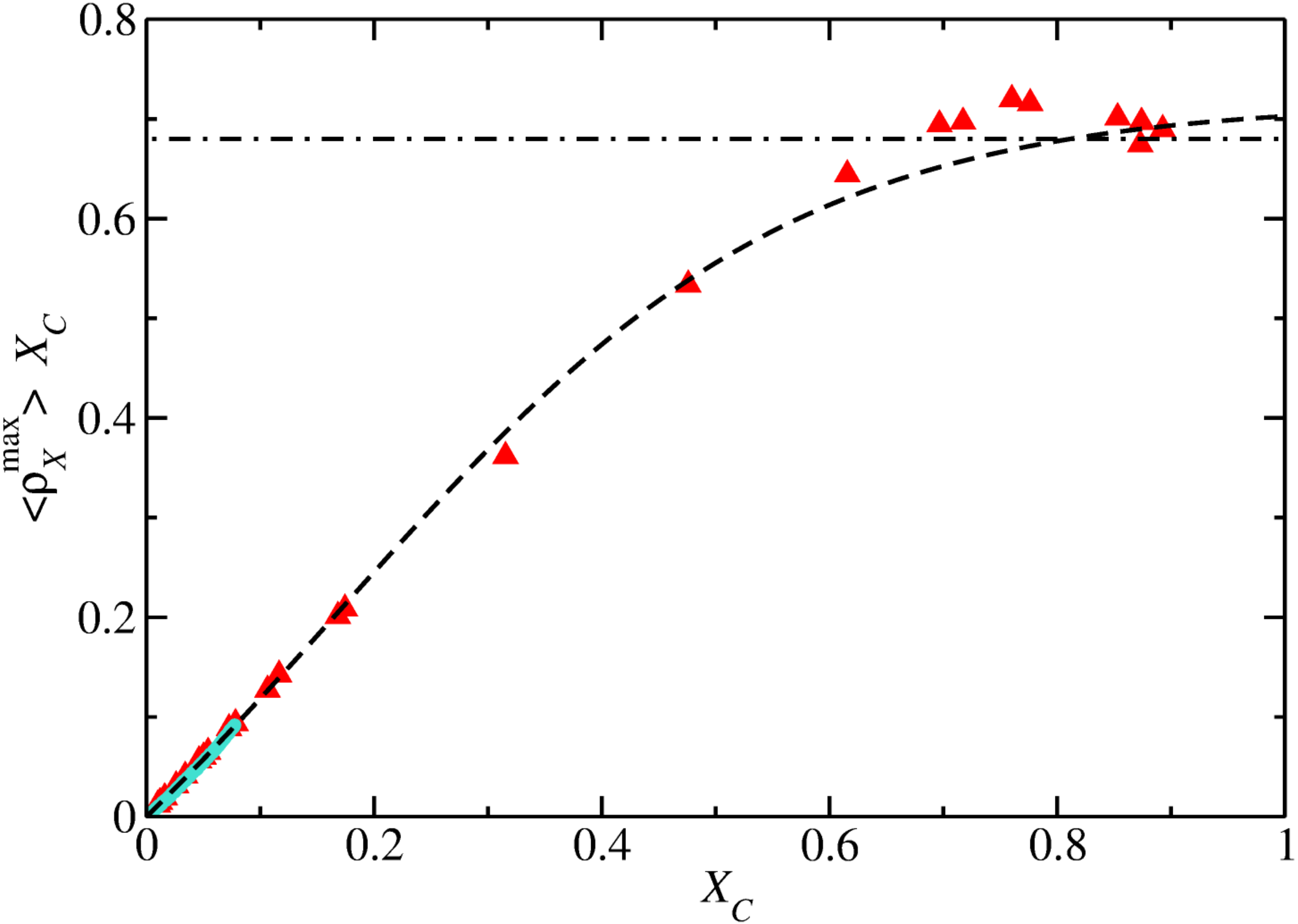}};
        \begin{scope}[x={(image.south east)},y={(image.north west)}]
            \node[anchor=south west,inner sep=0] (image) at (0.15,0.715) {\includegraphics[width=0.2\linewidth]{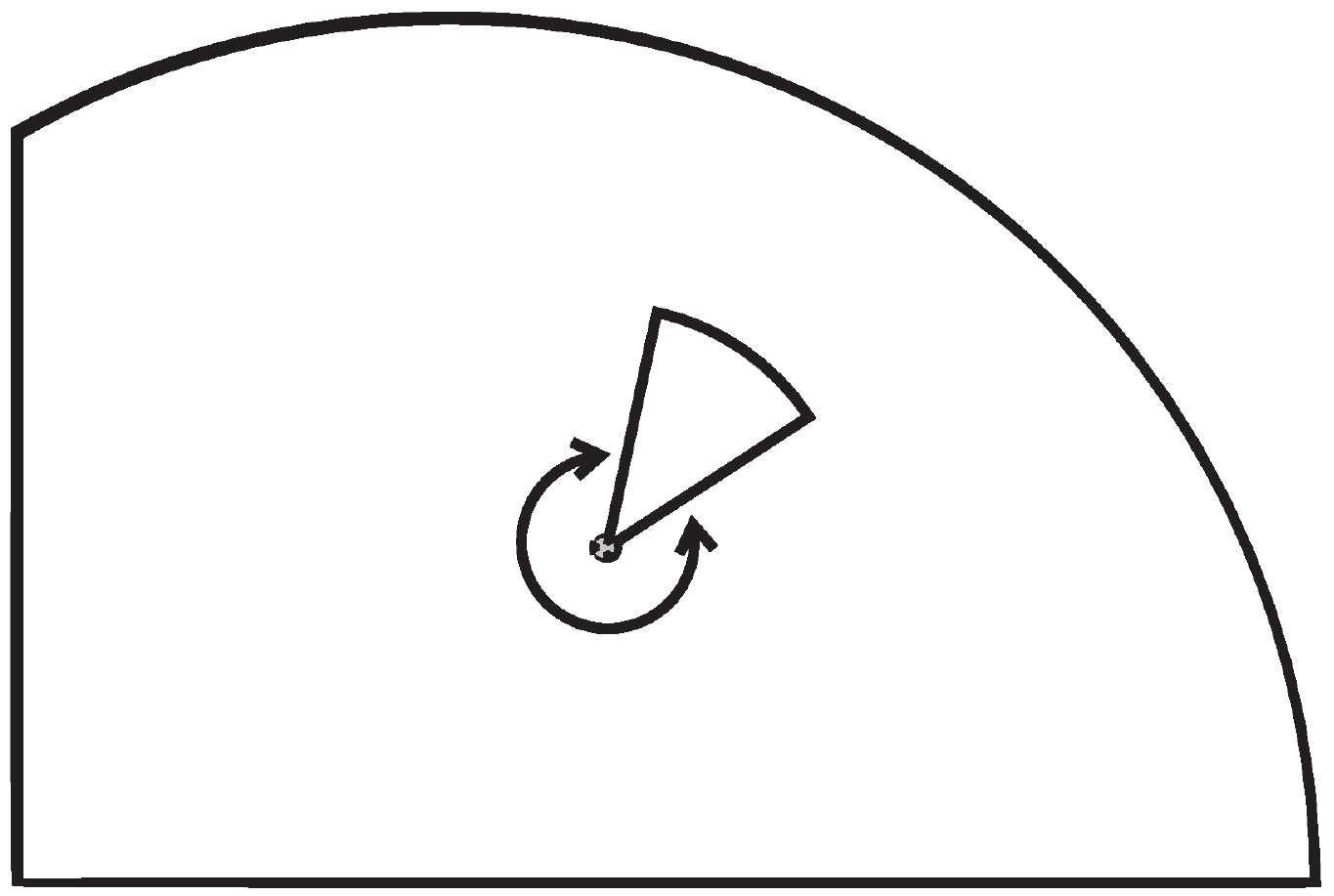}};
            \node[anchor=south west,inner sep=0] (image) at (0.54,0.162) {\includegraphics[width=0.35\linewidth]{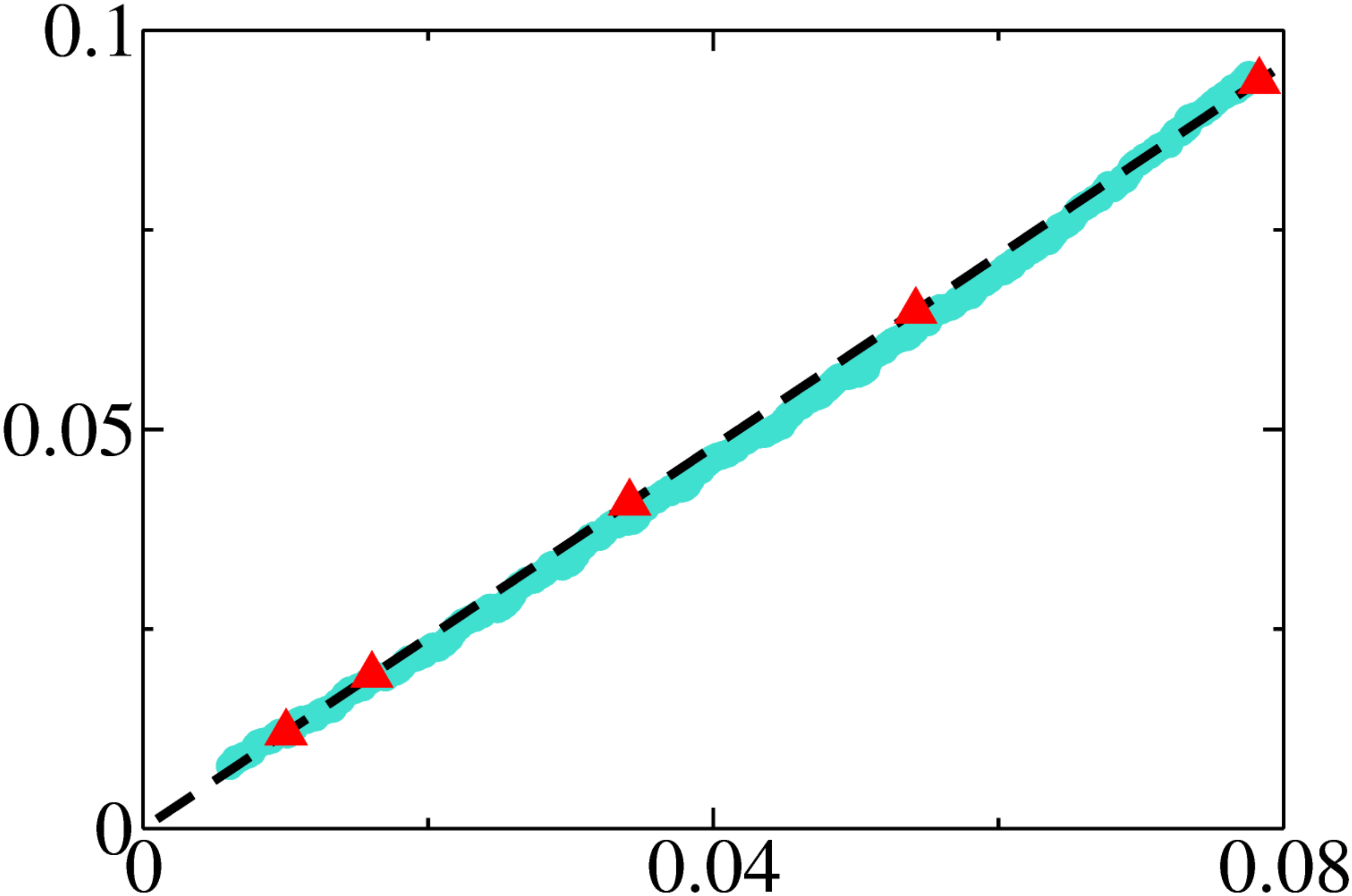}};
        \end{scope}
    \end{tikzpicture}
        \caption{Product of the mean density of maxima and the mean correlation width obtained for the parameter-dependent cut-circle system (turquoise dots) at $440$ different frequencies. The statistical errors are no larger than the size of the symbols. The dashed curve illustrates the analytical expression Eq.~(\ref{eqn:hussein2}) for the variation of an external parameter which describes the data well, with $c_0=1$, $c_1=-1/2$ and $c_2=1/3$. Upper inset: sketch of the billiard along with the rotatable wedge-shaped piece of Teflon$^{\circledR}$. The angle of rotation of the wedge defines the parameter but its initial orientation is arbitrary. Lower inset: zoom into the experimental data range $X_C\lesssim 0.08$.}
\label{fig7}
\end{figure}

The analysis of the fluctuation properties of the eigenvalues of closed, parameter dependent systems provided new insight into the universal behavior of chaotic systems~\cite{Gaspard1990,Simons1993,Guhr1998,Weidenmueller2005,Pato2005}. In order to study cross-section fluctuations of an open system with respect to a parameter, as opposed to the frequency, we analyzed the spectra measured with a microwave billiard having the shape of a desymmetrized straight-cut circle presented in Ref.~\cite{Dietz2006} and shown schematically in Fig.~\ref{fig7}, the classical dynamics of which is chaotic~\cite{Dietz2006,Ree1999}. The varying parameter, $\alpha$, concured with the angle of rotation of a dielectric wedge of Teflon$^{\circledR}$ inside the resonator and assumed 37 equally spaced values in steps of $2.5^\circ$. In order to ensure a sufficiently high spectral resolution and to eliminate dissipative processes, the experiment was performed at $4.2$~K with a superconducting microwave billiard manufactured from lead-plated copper~\cite{Richter1999}. Thus, complete sequences of 440 eigenfrequencies could be identified for each value of $\alpha$. The fluctuating positions of the resonance frequencies trace out irregular oscillatory curves that avoid intersections as the parameter $\alpha$ is varied (see Fig.~2 of Ref.~\cite{Dietz2006}). 

The eigenfrequencies were unfolded to mean spacing unity for a fixed value of $\alpha$ with the help of Weyl's formula, yielding 440 rescaled eigenfrequencies $e_i(\alpha),\, i=1,2,\cdots ,440$ for each of the 37 values of $\alpha$. Also the parameter $\alpha$ was rescaled by using the procedure described in Ref.~\cite{Leboeuf1999}, 
\begin{equation}
X=\sqrt{\langle v^2_\alpha\rangle}\cdot\alpha,\, \langle v^2_\alpha\rangle =\frac{1}{440}\sum_{i=1}^{440}\left(\frac{de_i}{d\alpha}\right)^2.
\end{equation}
We considered the cross-section autocorrelation function for parametric variations,
\begin {equation}
\mathcal{C}^{X}_{21} (\delta X) = \langle \lvert S_{21}(X) \rvert^2 \lvert S_{21}(X+\delta X) \rvert^2 \rangle - \langle \lvert S_{21} \rvert^2 \rangle^2,
\end{equation}   
where $\langle\ldots\rangle$ denotes the average over the parameter and intermediate frequencies between the respective eigenvalue and the neighboring ones. This correlation function has been examined for quantum dots and has a square Lorentzian shape~\cite{Guhr1998,Ramos2011,Barbosa2013}, $\mathcal{C}^{X}_{21} (\delta X)=\mathcal{C}^{X}_{21}(0)/\left(1+\left(\delta X/X_C\right)^2\right)^2$. The correlation width $X_C$ of $\mathcal{C}^{X}_{21}(\delta X)$ was either obtained from a fit of this formula to it or by determining the value of $\delta X$ for which $\mathcal{C}^{X}_{21}(\delta X)=\mathcal{C}^{X}_{21}(0)/4$. Both procedures yielded the same values of $X_C$. Maxima were counted in the spectrum $\lvert S_{21}(X)\rvert^2$ for each frequency, using the fact that the eigenfrequencies correspond to the frequencies at the maxima in the resonance spectra, $\lvert S_{21}(f)\rvert^2$. Accordingly, we determined the number of maxima in a given $X$ range by simply counting the eigenfrequencies, which actually had been determined for the studies published in Ref.~\cite{Dietz2006}, and the density of maxima, $\langle \rho^{\max}_{X} \rangle$, was inferred on dividing the number thereof by the range of the rescaled parameter. Figure~\ref{fig7} shows the quantity $\langle \rho^{\max}_X\rangle X_C$  as a function of the correlation width $X_C$. It is expected to assume a constant value of $3/\big(\pi\sqrt{2}\big)\simeq 0.68$ (dash-dotted line) for $X_C\gtrsim 1$. The absence of such saturation may be attributed to the attainment of only very small values of $X_C$, since the measurements were performed at superconducting conditions. Consequently, there is no dissipation in the walls, implying that the effective number of open channels is 2. This is in contrast to the experiments with the tilted-stadium microwave billiard at room temperature, where fictitious open channels had to be introduced to account for the absorption. For very small correlation widths $X_C$ an analytical expression, which is similar to that given in Eq.~(\ref{eqn:hussein1}) was derived in Ref.~\cite{Barbosa2013},

To test the results for the parametric system in more detail, we in addition performed RMT simulations using the $S$-matrix formalism Eq.~(\ref{eqn:Sab}). We replaced $H$ by a Hamiltonian $H(\mu)=H_1\cos\mu +H_2\sin\mu$, where $H_1,\, H_2$ are random matrices drawn from the GOE and $\mu$ is varied~\cite{Dietz2006}. The number of open channels was chosen equal to $M=32$. Still the results of these simulations, plotted as red triangles in Fig.~\ref{fig7}, describe the experimental data very well and saturate for $X_C\gtrsim 0.7$ at $\langle\rho^{\max}_{X}\rangle X_C\simeq 0.68$, i.e., at the value predicted for the Ericson region. Like in the experiments with the tilted stadium in~\refsec{Exp1} we derived an analytical expression for the dependence of the product $\langle\rho^{max}_X\rangle X_C$ on the correlation width $X_C$ in the regions of isolated and overlapping resonances using a similar ansatz
\begin{equation}
\langle \rho^{\max}_X\rangle X_C=\frac{3}{\pi\sqrt{2}}c_0\frac{X_C}{\sqrt{X_C^2+c_1X_C+c_2}},
\label{eqn:hussein2}
\end{equation}
with now 3 fit parameters $c_0,\, c_1,\, c_2$. The good agreement with Eq.~(\ref{eqn:hussein2}) (dashed line), obtained by setting $c_0=1,\, c_1=-1/2$, and $c_3=1/3$ is clearly visible. For large values of $X_C$, the RMT results for $\langle \rho^{\max}_X\rangle X_C$ and the analytical expression attain the predicted asymptotic value, $\langle \rho^{\max}_X\rangle X_C\to\frac{3}{\pi\sqrt{2}}$. The experimental data are also well described by this expression. This is further demonstrated in the inset which shows a zoom into the experimental data range.

In Refs.~\cite{Ramos2011,Hussein2014} an analytical expression was derived for the product of the mean density of maxima $\langle\rho^{\rm QD}_{X}\rangle$ and the width $X_C$ of the correlation function for the variation of an external perpendicular magnetic field in quantum dots, which takes the shape of a square Lorentzian in the Ericson region, 
\begin{equation}
\langle \rho^{QD}_{X}\rangle X_C =\frac{\sqrt{3}}{\pi\sqrt{2}}\sqrt{\frac{7\Gamma^2-10\Gamma+6}{2\Gamma^2-3\Gamma+2}},
\label{eqn:hussein2ex}
\end{equation}
with $\Gamma$ the tunneling probability, which corresponds to the average of the transmission coefficients $T_1$ and $T_2$ associated with the antennas in the microwave experiment. Figure~\ref{fig8} shows the ratio $\chi (\Gamma)=\langle \rho^{\max}_X\rangle X_C/(\langle\rho^{QD}_{X}\rangle X_C)$ of the experimental result and the analytical prediction Eq.~(\ref{eqn:hussein2ex}) for the product of the mean density of maxima and the mean correlation width. It illustrates that, like in the parameter-independent cases considered in~\refsec{Exp1}, the former are well described by the latter for $\Gamma\gtrsim 0.4$. While for quantum dots, the product of the mean density of maxima and the width of the corresponding correlation function attains a non-vanishing value for $\Gamma\to 0$, this is not the case for an open microwave or quantum billiard. Therefore, the deviation of $\chi(\Gamma)$ from unity, observed for small values of $\Gamma$, was expected.

\section{Conclusions} Drawing inspiration from ideas developed in nuclear physics, in this article the relation of the mean density of maxima $\langle \rho^{\max}_{\varepsilon}\rangle$ in cross-section fluctuations of a chaotic compound-nucleus scattering process to the correlation width $\Gamma_{\rm corr}$ using the analogy of the associated $S$-matrix formalism to that for the description of the resonance spectra of microwave billiards has been investigated. Historically, the value of $b_N=2K_N \Gamma_{\rm corr}$ has been extensively debated~\cite{Brink1963, Dallimore1965, Bizzeti1967} in the nuclear physics literature. Therefore it is reassuring that our experiments were able to reconfirm the prognostication of Refs.~\cite{Brink1963, Dallimore1965} regarding the saturation value for fairly large values of $\Gamma_{\rm corr}$ that were predicted nearly half a century ago. Not only do we establish the association thereof with conductance fluctuations in quantum dots but our numerical and RMT results also suggest an appropriate ansatz to account for the behavior in the region of low $\Gamma_{\rm corr}$. Indeed, on the basis of the fit of a general ansatz for the product of the mean density of maxima and the mean correlation width to the experimental and RMT results a good analytical description was found, which interpolates between zero for a vanishing correlation width and the predicted saturation value in the Ericson region, for the cases of a large number of open channels and of just 2 open channels and for a parameter dependent system, see Eqs.~(\ref{eqn:hussein1}),~(\ref{eqn:hussein11}) and~(\ref{eqn:hussein2}), respectively. Interestingly, all three expressions have the same simple structure and also the coefficients are similar. Still, we were not able to provide an analytical derivation. We also compared our experimental and numerical results with predictions for quantum dots~\cite{Hussein2014} associated with a correlation function with the shape of a Lorentzian and a squared Lorentzian, respectively, and found a good agreement for sufficiently large tunneling probabilities, i.e., transmission coefficients $\Gamma\gtrsim 0.4$. For small $\Gamma\to 0$, however, the product of the mean density of maxima and the mean correlation width vanishes in our experiments and RMT simulations and takes a non-zero value in quantum dots. The extension of the results to quantum graphs further broadens their scope of applicability to studies in condensed matter and solid-state physics for which quantum graphs have gained widespread acceptance as model systems~\cite{Montroll1970}. Hence, we believe that the results presented herein hold great potential at the interface of multiple fields for future investigations in this direction, while simultaneously bringing long-awaited closure to a lingering problem in nuclear physics.

\begin{acknowledgments}
This work was performed during an internship of RS at the Institute of Nuclear Physics of the Technical University of Darmstadt and supported by the DFG within the Collaborative Research Center 634. We acknowledge stimulating and clarifying discussions with Mahir Hussein. RS thanks for the kind hospitality during his stay. 
\end{acknowledgments}

%
%

\end{document}